\documentclass[epj]{svjour}

\newcommand{\beqa}{\begin{eqnarray}}
\newcommand{\eeqa}{\end{eqnarray}}
\newcommand{\ket} [1] {\vert #1 \rangle}
\newcommand{\bra} [1] {\langle #1 \vert} 

\newcommand{\proj}[1]{\ket{#1}\bra{#1}}
\newcommand{\mean}[1]{\langle #1 \rangle}

\usepackage[dvips]{graphicx}
\usepackage{amsmath}

\begin{document}

\bibliographystyle{amsplain}

\title{Cloning and Cryptography with Quantum Continuous Variables\thanks{Submitted to the special issue of the European Physical Journal D on ``Quantum interference and cryptographic keys: novel physics and advancing technologies''. Proceedings of the conference QUICK 2001, Cargese, Corsica, April 7--13, 2001.}}

\author{N. J. Cerf\inst{1,2} \and S. Iblisdir\inst{1} \and G. Van Assche\inst{1}}

\institute{Ecole Polytechnique, CP 165, Universit\'e Libre de Bruxelles, 1050 Brussels, Belgium \and Jet Propulsion Laboratory, California Institute of Technology, Pasadena, CA 91109, USA}

\date{July 2001}

\abstract{The cloning of quantum variables with continuous spectra is investigated. We define a Gaussian 1-to-2 cloning machine, which
copies equally well two conjugate variables such as position and momentum
or the two quadrature components of a light mode. The resulting cloning 
fidelity for coherent states, namely $F=2/3$, is shown to be optimal. An asymmetric version of this Gaussian cloner is then used to assess the security
of a continuous-variable quantum key distribution scheme that allows
two remote parties to share a Gaussian key. The
information versus disturbance tradeoff underlying this continuous quantum cryptographic scheme is then analyzed for the optimal individual attack. 
Methods to convert the resulting Gaussian keys into secret key bits are also studied.
The extension of the Gaussian cloner to optimal $N$-to-$M$ continuous 
cloners is then discussed, and it is shown how to implement these cloners for light modes, using a phase-insensitive optical amplifier and beam splitters. 
Finally, a phase-conjugated inputs $(N,N')$-to-$(M,M')$ continuous cloner
is defined, yielding $M$ clones and $M'$ anticlones from $N$ replicas
of a coherent state and $N'$ replicas of its phase-conjugate (with
$M'-M=N'-N$). This novel kind of cloners is shown to outperform the standard $N$-to-$M$ cloners in some situations.}
\PACS{{03.67.Dd}{} \and {03.65.Bz}{} \and {42.50.-p}{} \and {89.70.+c}{}}
\maketitle

\section{Introduction}

Quantum information theory was originally developed for discrete quantum variables, in particular quantum bits (qubits). Recently, however, it has been discovered that several concepts that were invented for qubits extend very naturally to the domain of continuous variables (e.g., position and momentum of a particle, or the quadrature components of a mode of the electromagnetic field). The first result in this direction concerned quantum teleportation \cite{vaid94,brau98:telep}, and gave rise to a lot of interest in continuous-variable quantum information processing.
In the present paper, we focus on the notions of quantum cloning and quantum key distribution, and investigate how they can be extended to continuous variables.

Cloning machines (that achieve the  optimal approximate cloning transformation compatible with the so-called no-cloning theorem) have been a fundamental research topic in the last five years. In Section~\ref{sec:qcm}, we will
define a Gaussian cloner, which achieves the optimal cloning of a continuous variable compatible with the requirement of being covariant with respect
to displacements and rotations in phase space. 
In other words, this cloner duplicates all coherent states with a same
fidelity ($F=2/3$). The optical implementation of this cloner and its
extension to $N$-to-$M$ cloners are also discussed. In Section~\ref{sec:qdgk},
we then turn to quantum key distribution, and propose a continuous-variable
cryptosystem that allows two remote parties to share a Gaussian key by
exchanging continuous key elements carried by squeezed states. This scheme
is the proper continuous counterpart of the protocol BB84 \cite{benn84} for qubits. Our continuous cryptosystem is related to the Gaussian cloner for an asymmetric version of the latter achieves the optimal individual eavesdropping strategy. Thus, our previous results on continuous cloning can be used to analyze the information versus disturbance tradeoff, in order to assess the security of this continuous cryptosystem. We find that the information gained by the eavesdropper is exactly upper bounded by the information lost by the authorized receiver. We also investigate a protocol to convert the raw Gaussian keys into a string of secret key bits, that is, we show how to apply reconciliation and privacy amplification on continuous key elements.
Finally, in Section~\ref{sec:pcic}, we come back to the issue of cloning
continuous variables, and define a new class of ``phase-conjugated inputs''
cloners. These cloners produce several clones (and anticlones) from
several replicas of an input coherent state {\em and} its phase conjugate. 
We show that adding these extra phase-conjugated inputs makes it possible to improve the cloning (and anticloning) fidelity with respect to the standard $N$-to-$M$ cloners. 

\section{Quantum Cloning Machines}
\label{sec:qcm}

Let us first seek for a transformation which duplicates with a same fidelity all coherent states $\ket{\psi}$, with $\psi=(x+ip)/\sqrt{2}$.  The fundamental requirement we put on this transformation is that it is covariant with respect to displacements in phase space. That is, if two input states are identical up to a displacement $\hat D(x,p)=e^{-ix\hat p} e^{ip\hat x}$, then their respective copies should be identical up to the same displacement. (In this paper, we put $\hbar=1$). Thus, denoting by $\mathcal{H}$ the Hilbert space corresponding to a single copy, cloning can be defined as a completely positive trace-preserving linear map ${\mathcal{C}}: \mathcal{H} \to \mathcal{H}^{\otimes 2}: \proj{\psi} \to {\mathcal{C}} (\proj{\psi})$ such that 
\begin{equation}
\begin{split}
\label{eq:cov}
{\mathcal{C}} ( \hat{D}(x,p)\proj{\psi} \hat{D}^{\dagger}(x,p)) \\
= \hat{D}^{\otimes 2}(x,p) {\mathcal{C}} (\proj{\psi}) \hat{D}^{\dagger\otimes 2}(x,p),
\\
\end{split}
\end{equation}
for all displacements $\hat D(x,p)$ in the phase space. A simple way to meet displacement covariance is to seek for a cloning transformation whose output clone individual states are given each by a Gaussian mixture:
\begin{equation}
\begin{split}
\label{eq:gaussmix}
\rho(\proj{\psi})=\frac{1}{2\pi\sigma^2} 
\int dx \; dp  \; e^{-\frac{x^2+p^2}{2\sigma^2}}
\\
\times \hat D(x,p)\ket{\psi}\bra{\psi} \hat D^{\dagger}(x,p)  ,  
\end{split}
\end{equation}
where $\sigma^2$ is the cloning-induced error variance. In the following we will refer to such a transformation as a Gaussian cloner. Note that Eq.~(\ref{eq:gaussmix}) is such that the cloning induced noise on the quadratures $\hat{x}$ or $\hat{p}$ is invariant under rotations in the phase space, which is certainly a desirable property since it is satisfied by coherent states. Consider the following unitary operator:
\begin{equation}
\label{unitop}
\hat{U}_{1,2,3}=e^{-i(\hat{x}_3-\hat{x}_2)\hat{p}_1} e^{-i\hat{x}_1(\hat{p}_2+\hat{p}_3)}e^{-i\hat{x}_2\hat{p}_3},  
\end{equation}
where modes $1$, $2$ and $3$ refer respectively to the original, the additional copy, and an auxiliary mode (also refered to as an ancilla). This operator can be used to build a Gaussian cloner if the additional copy and the ancilla are initially prepared in the vacuum state\cite{cerf00:cont}. Indeed, it is readily checked that this transformation outputs two clones whose individual states are Gaussian-distributed, as in Eq.(\ref{eq:gaussmix}), with a variance $\sigma^2=1/2$. In particular, it copies all coherent states $\ket{\psi}$ with the same fidelity $f_{1,2}=\bra{\psi}\rho(\psi)\ket{\psi}=2/3$. 

This machine is optimal in the sense that it is impossible to have $\sigma^2(1,2) < 1/2$. To prove this, let us consider the following situation. A coherent state is processed through such a cloner, the observable $\hat{x}$ being measured at one output clone while the observable $\hat{p}$ is measured at the other output. Let us denote by $\Sigma^2_x$ and $\Sigma^2_p$ the respective error variances corresponding to this joint measurement. From the general theory on the simultaneous measurement of conjugate observables \cite{arth65}, we know that
\begin{equation}
\Sigma^2_x \Sigma^2_p \geq 1.
\end{equation}
Using Eq.~(\ref{eq:gaussmix}), we get 
\begin{equation}
(\delta \hat{x}^2+\sigma^2)(\delta \hat{p}^2+\sigma^2) \geq 1,
\end{equation}
where $\delta \hat{x}^2 (\delta \hat{p}^2)$ is the intrinsic variance of $\hat{x}$ ($\hat{p}$) of the input state and $\sigma^2$ is the cloning-induced variance. Now, using the uncertainty principle $\delta \hat{x}^2 \delta \hat{p}^2 \geq 1/4$ and the inequality $a^2+b^2 \geq 2 \sqrt{a^2 b^2}$, we conclude that $\sigma^2 \geq 1/2$, implying that the unitary operator Eq.~(\ref{unitop}) is indeed optimal to achieve Gaussian cloning\cite{cerf00:coherent}.
 
\begin{figure}
\includegraphics[width=3.0in,angle=0]{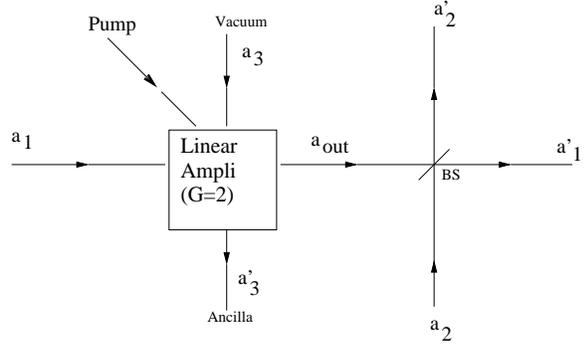}
\caption{Implementation of a $1 \to 2$ cloner using a phase-insensitive linear amplifier and a $50:50$ beam-splitter (BS).}
\label{fig:ccircuit2}
\end{figure}

A possible implementation of this machine (see Fig. \ref{fig:ccircuit2}) consists in processing the input mode into a linear phase-insensitive amplifier \cite{cave82} of gain $G=2$:
\begin{equation}
\hat{a}_{out}=\sqrt{2}\hat{a}_{1}+\hat{a}_{3}^{\dagger}, \qquad 
\hat{a'}_{3}=\hat{a}_{1}^{\dagger}+\sqrt{2}\hat{a}_{3},
\end{equation}
with $a_j=(\hat{x_j}+i\hat{p_j})/\sqrt{2}$ denoting the annihilation operator for mode $j$). Then, one produces the two output clones by processing the output signal of the amplifier through a $50:50$ phase-free beam-splitter:
\begin{equation}
\hat{a'}_{1}=\frac{1}{\sqrt{2}}(\hat{a}_{1}+\hat{a}_{2}), \qquad 
\hat{a'}_{2}=\frac{1}{\sqrt{2}}(\hat{a}_{1}-\hat{a}_{2}),
\end{equation}
It is readily checked that this scheme leads to an equal $x$-error and $p$-error variance of $1/2$ for both clones, that is, it achieves the optimal Gaussian cloner.

We will now present two generalizations of this $1 \to 2$ Gaussian quantum cloning machine. The first one consists in a transformation which from $N$ ($\geq 1$) original input states produces $M$ ($\geq 2$) output copies whose individual states are again given by an expression similar to Eq.(\ref{eq:gaussmix}), but with a different error variance ${\sigma^2}_{N,M}$. Using an argument based on the concatenation of cloners, it is possible to derive a lower bound on ${\sigma^2}_{N,M}$, that is\cite{cerf00:coherent}
\begin{equation}
\label{eq:varclon}
{\sigma^2}_{N,M} \geq \frac{1}{N}-\frac{1}{M},
\end{equation}
with the corresponding fidelity for coherent states
\begin{equation}
\label{eq:fidclon}
f_{N,M} \leq \frac{MN}{MN+M-N}.
\end{equation}
Again, these bounds can be attained by a transformation whose implementation necessitates only a phase-insensitive linear amplifier and beam splitters \cite{brau00}. Loosely speaking, the procedure consists in concentrating the $N$ input modes into a single mode by a network of beam splitters, to amplify the resulting mode, and then to distribute the output amplified mode into $M$ output modes through a second network of beam-splitters. Note that the bounds Eq.~(\ref{eq:varclon},\ref{eq:fidclon}) can also be derived using techniques similar to the ones used for evaluating quantum nondemolition measurements. This was been done in a paper establishing a link between cloning and teleportation for continuous variables \cite{gros01}.

The second generalization of the $1 \to 2$ Gaussian quantum cloning machine we will briefly consider here is the case where the $\hat{x}$ and $\hat{p}$ quadratures are not treated equally, and the case where the two output clones do not have the same fidelity. Equation~(\ref{eq:gaussmix}) then has to be replaced by
\begin{equation}
\begin{split}
\label{eq:gaussmixasym}
\rho(\proj{\psi})=\frac{1}{2\pi\sqrt{\sigma^2_{i,x}\sigma^2_{i,p}}} 
\int dx \; dp  \; e^{-(\frac{x^2}{2\sigma^2_{i,x}}+\frac{p^2}{2\sigma^2{i,p}})} \\ 
\times \hat D(x,p)\ket{\psi}\bra{\psi} \hat D^{\dagger}(x,p),  
\end{split}
\end{equation}
where $\sigma^2_{i,x}$ (resp. $\sigma^2_{i,p}$) stands for the cloning-induced error variance in the quadrature $\hat{x}$ (resp. $\hat{p}$) for the $i$th clone. In this case, it is possible to prove \cite{cerf00:cont} that the following cloning uncertainty relations must hold:
\beqa   \label{eq_uncert}
\sigma^2_{1,x} \sigma^2_{2,p} &\ge& 1/4,  \nonumber\\
\sigma^2_{1,p} \sigma^2_{2,x} &\ge& 1/4.
\eeqa

Asymmetries between the output clones and between the $x$/$p$ variables can be characterized by the following two parameters:
\begin{equation}
\chi = \frac{\sigma_{1,x}}{\sigma_{2,x}} = \frac{\sigma_{1,p}}{\sigma_{2,p}}, \mbox{ and }
\lambda = \frac{\sigma_{1,x}}{\sigma_{1,p}} = \frac{\sigma_{2,x}}{\sigma_{2,p}}.
\end{equation}

As suggested in \cite{fiur01}, asymmetric machines (with $\chi\neq 1$) can be implemented by a scheme akin to Fig.~\ref{fig:ccircuit2} in the sense that only two beam splitters and a single linear amplifier are needed.
We will see in the following section how these asymmetric quantum cloning machines can be used to assess the security of a continuous-variable quantum key distribution protocol.

\section{Quantum Key Distribution}
\label{sec:qdgk}

In this section, we introduce a quantum protocol for the distribution of Gaussian key elements, which is a continuous-variable analogue of 
the protocol BB84 \cite{benn84} -- we assume here that the reader is familiar with BB84. Our protocol, introduced in \cite{cerf00:qdgk}, works like BB84 but with binary information being replaced by continuous information that behaves essentially like in a Gaussian channel. 

One exploits a pair of canonically conjugate continuous variables $x$ and $p$, which can be thought of, for instance, as the two quadratures $X_1$ and $X_2$ of the amplitude of a mode of the electromagnetic field \cite{scully}. Alice randomly chooses a random key element $r$ that follows a Gaussian distribution with mean zero and variance $\Sigma^2$, and randomly decides to encode it into either $x$ (i.e., $\mean{x}=r$) or $p$ (i.e., $\mean{p}=r$). An eavesdropper ignoring which of these two encoding rules is used cannot acquire information without disturbing the state.

Let us now describe the exact nature of the states used for encoding each key element. When encoding the value $r \sim N(0, \Sigma_x)$ in $x$, Alice creates a Gaussian state such that $\mean{x}=r$, $\mean{p}=0$, $\Delta x^2 = \sigma^2_x$ and thus $\Delta p^2 = 1/4\sigma^2_x$. Similarly, when the value $r \sim N(0, \Sigma_p)$ is encoded in $p$, the encoding state has $\mean{p}=r$, $\mean{x}=0$, $\Delta p^2 = \sigma^2_p$ and thus $\Delta x^2 = 1/4\sigma^2_p$.

On his side, Bob measures either $x$ or $p$ at random. Like in BB84, half of the measurements give results that are uncorrelated to Alice's values, so half of the samples must be discarded when Alice discloses the encoding variable. Unlike BB84, however, measuring the correct variable does not yield the exact value of $r$, even with a perfect apparatus, because of the intrinsic noise of the Gaussian state. The value $r$ follows a Gaussian distribution $N(0, \Sigma_{x,p})$, to which some Gaussian noise is added $N(0, \sigma_{x,p})$, thus resulting in a Gaussian distribution with variance $\Sigma^2_{x,p}+\sigma^2_{x,p}$. We can therefore model the transmission of $r$ as a Gaussian channel with a signal-to-noise ratio (SNR) equal to $\Sigma_{x}^2 / \sigma_{x}^2$ or $\Sigma_{p}^2 / \sigma_{p}^2$.

An important requirement of the protocol is to make it impossible for Eve to be able to infer which encoding variable Alice used. For this, measuring the correct or incorrect variable ($x$ or $p$) must yield statistically indistinguishable results. If, in contrast, Eve was able to detect (even not perfectly) that she measured the wrong set, then she could fake an attenuation by discarding wrong key elements and retransmitting only the correctly measured ones. This indistinguishability requirement can be expressed as the equality of the density matrices resulting from the two encoding rules, or equivalently as \cite{cerf00:qdgk}
\begin{equation}
\label{eq:variances}
1+\frac{\Sigma_x^2}{\sigma_x^2} = 1+\frac{\Sigma_p^2}{\sigma_p^2} = \frac{1}{4\sigma_x^2 \sigma_p^2}.
\end{equation}
A proof of this is given in Appendix~\ref{sec:density}. This also means that the SNR is the same for both variables $x$ and $p$, and that the information rate is \cite{cover}
\begin{equation}
I = \frac{1}{2} \log_2(1+\Sigma_x^2 / \sigma_x^2) = -\log_2 (2 \sigma_x \sigma_p).
\end{equation}

\subsection{Eavesdropping by cloning}

Let us now discuss an individual eavesdropping of this protocol with cloning machines such as those defined in Section~\ref{sec:qcm}. Eve makes two clones of the state sent by Alice, one of which is transmitted to Bob, and the other is measured in the correct variable when Alice reveals the encoding rule. This happens to be the optimal individual eavesdropping strategy as shown in \cite{cerf00:qdgk} and \cite{ralp00:security}.

We use a $1 \to 2$ cloning machine, and we keep the freedom to make a better clone for Bob or Eve (parameter $\chi$) and to get more accuracy in $x$ or $p$ (parameter $\lambda$). The subscripts $1$ and $2$ for the two copies are replaced respectively by $B$ and $E$ for the two recipients. The added variances on the clones will be:
\beqa
&\sigma^2_{B,x} = \frac{1}{2}\chi \lambda, \quad &\sigma^2_{B,p} = \frac{1}{2}\chi \lambda^{-1}, \\
&\sigma^2_{E,x} = \frac{1}{2}\chi^{-1} \lambda, &\sigma^2_{E,p} = \frac{1}{2}\chi^{-1} \lambda^{-1}.
\eeqa

Let us calculate the resulting information rates. When Bob measures $x$, the result is affected both by the intrinsic fluctuations of $x$ and by the noise induced by the cloning operation, thus resulting in a total variance $\sigma^2_x + \frac{1}{2}\chi \lambda$. This is the noise power in the Gaussian channel representing the communication between Alice and Bob through Eve's cloning machine. Therefore, the information rate is now
\begin{equation}
I_{B,x} = \frac{1}{2} \log_2 (1+\frac{\Sigma^2_x}{\sigma^2_x + \frac{1}{2}\chi \lambda}).
\end{equation}

Similarly, one can calculate the new variance on $p$ measured by Eve on her clone, namely $\sigma^2_p + \frac{1}{2} \chi^{-1} \lambda^{-1}$. This gives an information rate
\begin{equation}
I_{E,p} = \frac{1}{2} \log_2 (1+\frac{\Sigma^2_p}{\sigma^2_p + \frac{1}{2}\chi^{-1} \lambda^{-1}}).
\end{equation}

Adding the last two information rates indicates the balance between Bob's and Eve's information. Remarkably, the information that Eve gains by using this attack on $p$ is exactly equal to the information that Bob loses on $x$ \cite{cerf00:qdgk},
\begin{equation}
I_{B,x}+I_{E,p} = \frac{1}{2} \log_2 (1+\frac{\Sigma^2_x}{\sigma^2_x})=I.
\end{equation}
Of course, this balance also works when swapping $x$ and $p$, namely $I_{B,p}+I_{E,x} = I$.

This result is interesting because it allows Bob to bound from above the information gained by a possible eavesdropper. Assuming symmetry of the protocol in $x$ and $p$, Bob can estimate $I-I_B$ and is guaranteed that $I_E \leq I - I_B$ (in practice, a part of the information loss will be due to channel noise). From Ref.~\cite{maur93}, it is kwown that with reconciliation and privacy amplification carried out over a public authenticated channel, one is guaranteed to generate key bits whenever $I_B > I_E$. This last condition is in turn guaranteed provided that $I_B > I/2$, so that up to a 50\% information loss on Bob's side is acceptable in order to generate key bits.

\subsection{From Gaussian key elements to secret bits}

Let us now investigate the classical part of the key distribution protocol since we have to deal with reconciliation and privacy amplification based on \emph{continuous} raw key elements here, in contrast to BB84.
Shannon's formula gives us an upper limit on the number of bits one can send through a Gaussian channel with a given SNR. In our protocol, neither Alice nor Bob chooses the Gaussian random values. Yet, we want them to be able to extract a common string of bits out of their correlated Gaussian values, revealing as little information as possible on the public channel.

Our secret key distillation procedure \cite{vana01} works in the following way. First, Alice and Bob are going to extract common bits out of their Gaussian-distributed values, using a binary correction algorithm such as Cascade or a variant \cite{bras93,sugi00,yama00,chen01}. They will use it several times, on several real-to-binary conversion functions. Then, the resulting bits will undergo the usual privacy amplification procedure \cite{maur93,benn88:pa,benn95:pa}, for instance using a universal class of hash functions.

Let $X$ denote the random variable representing Alice's Gaussian values, and $X'$ Bob's values. Alice uses a set of real-to-binary conversion functions $S_i(X)=0,1$, ($1 \leq i \leq m$). These are called \emph{slices}, in the sense that instead of performing reconciliation on the real-valued string $x_{1\dots l}$, we operate on each string $S_i(x_{1\dots l})$ sequentially, like slices of the main, real-valued string. On his side, Bob uses another set of functions $\tilde{S}_i$, called \emph{slice estimators}, which reflects his best guess on the bit $S_i(X)$ given his current knowledge. The slice estimator $\tilde{S}_i$ is not only a function of $X'$ but also of the previous slices, $\tilde{S}_i(X', S_1(X), \dots, S_{i-1}(X))$. This results from the fact that the slices are corrected sequentially for $i=1, \dots, m$, and thus upon correcting slice $i$ Bob already knows $S_1(X), \dots S_{i-1}(X)$.
By carefully choosing the functions $S_i$ and $\tilde{S}_i$, both parties can extract a common string of bits out of the correlated Gaussian values, while only disclosing a little more than $H(S_1(X),\dots,S_m(X)|X')$ bits on the public channel. A more detailed analysis is given in \cite{vana01}.

Let us take an example. Assume the channel has $\Sigma^2/\sigma^2=15$, which means that Alice and Bob can share up to $I=\frac{1}{2}\log_2(1+\Sigma^2/\sigma^2)=2$ bits per raw key element. We assume $m=5$ slices as a trade-off between the efficiency of large $m$ and the use of reasonable computing resources. The slice functions $S_i(X)$, $1 \leq i \leq 5$ are constructed in the following way. First, the Gaussian distribution of $X$ is divided into $2^m=32$ intervals. The interval labeling function $T(X)$, which associates an interval number (from 0 to 31) to each value of $x$, is chosen so as to maximize $I(T(X);X')$. Thus, Bob starts with an optimal knowledge of $T(X)$. Then, we create the slice functions by assigning bit values to each of these intervals. Stated otherwise, we create a bijection between $S_{1\dots 5}(X)$ and $T(X)$ so that each vector of the 5 slice bits represents one (and only one) interval defined by $T(X)$. Much freedom is permitted at this step, but what we found to work best is to assign the least significant bit of the interval number to $S_1(X)$, the second bit to $S_2(X)$, and so on up to the most significant bit to $S_5(X)$.

The slice estimator functions $\tilde{S}_{1\dots 5}(X', \dots)$ are constructed from the slices $S_{1\dots 5}(X)$ and from the joint probability density $f_{X,X'}(x,x')$. Each estimator $\tilde S_i$ evaluates whether 
$S_i(X)=0$ or $S_i(X)=1$ is more likely conditionally on the arguments given to the estimator, namely $X'$ and the previous slices $S_{j<i}(X)$.

In the present example, Alice's and Bob's bits are almost uncorrelated when correcting slices 1 and 2.  The binary correction algorithm does not have to be used at this point -- it is enough for Alice to entirely reveal $S_1(X)$ and $S_2(X)$ for the whole string. Then, slice 3 on Alice's side and the slice estimator 3 on Bob's side produce two bit strings that match 76\% of the time -- it is thus possible to proceed with error correction using a binary correction algorithm. Note that the bit strings would be less correlated if the knowledge of $S_1(X)$ and $S_2(X)$ was not brought to Bob. Then for slice 4 (resp. slice 5), Alice's and Bob's string match 98\% (resp. 99.999\%) of the time, for which the binary correction will disclose only a small amount of information. Again, the knowledge of slices 1-3 helped Bob accurately estimate slice 4, which in turn helped him estimate slice 5.

As a result of this 5-step correction, Alice and Bob share a string of bits whose entropy is $H(S_{1\dots 5})=4.8$ bits per raw key element. Assuming a perfect binary correction algorithm, about 3 bits per raw key elements were disclosed. Roughly speaking, the net effect is thus $4.8-3=1.8$ bit of secret information per raw key element after privacy amplification (which is to be compared with the 2 bits per key element as given by Shannon's formula).

This is of course only an example. More elaborate constructions can be performed, such as gathering $d$ Gaussian key elements at once. In fact, it was shown in \cite{vana01} that the disclosed information reaches the Shannon bound as $d \to \infty$, just like for instance data compression works best for asymptotically large block sizes.

Now that we showed how quantum cryptography (followed by reconciliation and privacy amplification) can work with continuous variables, let us investigate another application of continuous variables to a special kind of quantum cloning machines.

\section{Phase-Conjugated Inputs Quantum Cloning Machines}
\label{sec:pcic}

\begin{figure}
\includegraphics[width=3.0in]{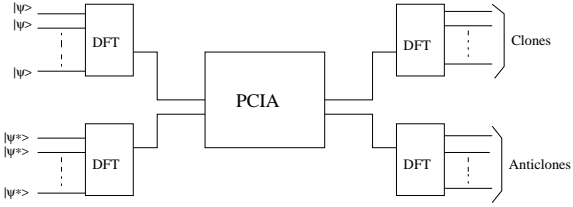}
\caption{PCI cloner that produces $M$ clones and $M'$ anti-clones
from $N$ replicas of $\ket{\psi}$ and $N'$ replicas of $\ket{\psi^*}$. Modes are concentrated and distributed by Discrete Fourier Transform (DFT).} 
\label{fig:pcic}
\end{figure}

It has been shown that an antiparallel pair of qubits is intrinsically more informative than a pair of parallel qubits if the goal is to encode a direction in space \cite{gipo99}.  Similarly for quantum continuous variables, one can show that more information can be encoded in a pair of phase- conjugated coherent states $\ket{\psi}\ket{\psi^*}$ than in two identical replicas $\ket{\psi}\ket{\psi}$ \cite{cerf01:pc}. Following on these ideas, we present here a phase-conjugated input (PCI) quantum cloning machine, that is, a transformation which taking as input $N$ replicas of a coherent state $\ket{\psi}$ and $N'$ replicas of its complex conjugate $\ket{\psi^*}$, produces $M$ optimal clones of $\ket{\psi}$ \cite{cerf01:pcic}. Again we will require that all the clones are treated equally, and that the cloner is covariant with respect to both displacements and rotations in phase space. As a matter of fact, it turns out that such a transformation can be implemented optimally using a sequence of beam-splitters, a single non-linear medium, and another sequence of beam-splitters, just as in the case of standard cloning. The procedure is the following (see Fig.~\ref{fig:pcic}):

(i) Concentrate the $N$ replicas of $\ket{\psi}$ stored in the $N$
modes $\{c_l\}$ ($l=0 \ldots N-1$) into a single mode $a_1$,
resulting in a coherent state of amplitude $\sqrt{N}\, \psi$. This
operation can be performed with a network of beam-splitters
achieving a $N$-mode Discrete Fourier Transform (DFT)\cite{brau00}. We get:
\begin{equation}
a_1=\frac{1}{\sqrt{N}}\sum_{l=0}^{N-1} c_l,
\end{equation}
and $N-1$ vacuum modes. Similarly, with a $N'$-mode DFT, concentrate the $N'$ replicas of $\ket{\psi^*}$ stored in the $N'$ modes $\{d_l \}$ $(l=0 \ldots N'-1)$ into a single mode $a_2$. This results in a coherent state of amplitude
$\sqrt{N'}\, \psi^*$. We have:
\begin{equation}
a_2=\frac{1}{\sqrt{N}}\sum_{l=0}^{N'-1} d_l.
\end{equation}

(ii) Apply the following transformation on the modes $a_1$ and $a_2$, resulting in modes $b_1$ and $b_2$ defined by

\beqa\label{amplifier}
b_1 &=& \sqrt{G} a_1 +\sqrt{G-1} a_2^{\dagger}, \nonumber \\
b_2 &=& \sqrt{G-1} a_1^{\dagger}+\sqrt{G} a_2,
\eeqa
where
\begin{equation}
\label{eq:gain}
\sqrt{G}=\frac{\sqrt{N' M'}-\sqrt{NM}}{N'-N},
\end{equation}
with
\begin{equation}
\label{eq:balance}
M'-M=N'-N.
\end{equation}
For obvious reasons, we call this transformation a 'phase-conjugated input amplification' (PCIA).

(iii) Distribute the output mode $b_1$ into $M$ clones $\{c'_l\}$ $(l=0
\ldots M-1)$ with a $M$-mode DFT:
\begin{equation}
\label{eq:dft}
c'_l=\frac{1}{\sqrt{M}}(b_1+e^{i \pi kl/M} v_k),
\end{equation}
where $\{v_k \}$ $(k=1 \ldots M-1)$ denote $M-1$ additional vacuum modes. It is readily verified that this procedure yields $M$ clones of $\ket{\psi}$. Interestingly, the amplitude $b_2$ of the other output of the PCIA has a mean value $\sqrt{M'}\psi^*$. Therefore, it can be used to produce $M'$ phase-conjugated clones (or anti-clones) of $\ket{\psi}$, $\{d'_l\}$ $(l=0 \ldots M'-1)$, using a $M'$-mode DFT:
\begin{equation}
\label{eq:dft2}
d'_l=\frac{1}{\sqrt{M'}}(b_2+e^{i \pi kl/M'} w_k)
\end{equation}
where $\{w_k\}$ $(k=1 \ldots M'-1)$ denote $M'-1$ additional vacuum modes. 

Some algebra shows that this procedure is optimal to produce $M$ clones, and that the additional $M'$ anti-clones are also optimal\cite{cerf01:pcic}. Furthermore, since the step (ii) of our procedure is linear and phase-insensitive, the resulting PCI cloner is covariant with respect to translations and rotations of the state to be copied: all coherent states are copied equally well, and the cloning-induced noise is the same for all quadratures. 

It is straightforward to calculate the noise of the clones and anti-clones: 
\beqa\label{eq:clonoise}
(\Delta {c'_l})^2 &=& \frac{1}{2}\mean{c'_l {c'_l}^{\dagger}+ {c'_l}^{\dagger} c'_l}-\mean{c'_l}\mean{c'_l}=\frac{1}{2}+\frac{G-1}{M} ,  \nonumber \\
(\Delta {d'_l})^2 &=&\frac{1}{2}+\frac{G-1}{M'} ,
\eeqa 
As expected, the variance of the output clones exceeds $1/2$, reflecting that perfect cloning (anti-cloning) is indeed impossible. Instead, they suffer from a thermal noise with a mean number of photons given by $\mean{n_{th}}=(G-1)/M$. In other words,
their $P$-function \cite{scully} is a Gaussian distribution
\begin{equation}
\label{eq:pfunc}
P(\xi,\xi^*)=\frac{1}{\pi \mean{n_{th}}} \; e^{-|\xi-\psi|^2/\mean{n_{th}}}.
\end{equation}
rather than a Dirac distribution $P(\xi,\xi^*)=\delta^{(2)}(\xi-\psi)$.

\subsection{Balanced phase-conjugated inputs cloner}

Consider now the balanced case ($N=N'$, $M=M'$), for which simple expressions of the noise variance and the fidelity can be obtained. We then have $G=(M+N)^2 / 4MN$, giving
\begin{equation}
\label{eq:pcinoise}
(\Delta {c'_l})^2=(\Delta {d'_l})^2= \frac{1}{2}+\frac{(M-N)^2}{4M^2N}.
\end{equation}
and
\begin{equation}
\label{eq:balfid}
f^N_{N, M}=\frac{1}{1+\mean{n_{th}}}=\frac{4M^2 N}{4M^2 N+(M-N)^2}.
\end{equation}

Let us compare these quantities to the variance and fidelity of a $2N \to M$ usual cloning machine, as obtained by replacing $N$ into $2N$ in Eqs. (\ref{eq:varclon}) and Eq.(\ref{eq:fidclon}). Of course, in the trivial case where $M=2N$, standard cloning can be achieved perfectly, while the balanced PCI cloner yields an additional variance $\mean{n_{th}}=1/(16N)$. However, whenever $M \geq 2N+1$, the $({}^{N}_{N})\to M$ PCI cloner outperforms the standard $2N \to M$ cloning machine. Also, comparatively more anti-clones with a higher fidelity are produced with the PCI cloner. Indeed, a standard $2N\to M$ cloning machine produces $M-2N$ anti-clones of fidelity $2N/2N+1$, which is actually the fidelity of an optimal measurement of $2N$ replicas of $\ket{\psi}$, whereas a PCI cloner produces $M$ anti-clones with a higher fidelity, as given by Eq.~(\ref{eq:balfid}). In particular, for $M \to \infty$, we see from Eq.~(\ref{eq:pcinoise}) that the additional noise induced by a PCI cloner is $1/4N$, that is, {\em one half} of the noise induced by a standard $2N \to \infty$ cloner (i. e., $1/2N$). In this case, the output of the PCIA can be considered as classical and the underlying process appears to be equivalent to a measurement. This reflects that more classical information can be encoded in $N$ pairs of phase-conjugated replicas of a coherent state rather than in $2N$ identical replicas, a result which was proven for $N=1$ in \cite{cerf01:pc}. More generally, in the unbalanced case ($N\ne N'$), it is readily checked, using Eq.(\ref{eq:gain}), that the optimal measurement results in a noise that is equal to that obtained by measuring $(\sqrt{N}+\sqrt{N'})^2$ identical replicas of the input, in the absence of phase-conjugated inputs.

\subsection{Unbalanced phase-conjugated inputs cloner}

As we have just shown, the balanced PCI cloner results in better cloning fidelities than a standard cloner. More generally, we may ask the following question: in order to produce $M$ clones of a coherent state $\ket{\psi}$ from a fixed total number $n$ of input modes, $N$ of which being in the coherent state $\ket{\psi}$ and $N'$ of which being in the phase-conjugated state $\ket{\psi^*}$, what is the phase-conjugate fraction $a=N'/n$ that minimizes the error variances of the clones? 

\begin{figure}
\includegraphics[angle=-90,width=3.0in]{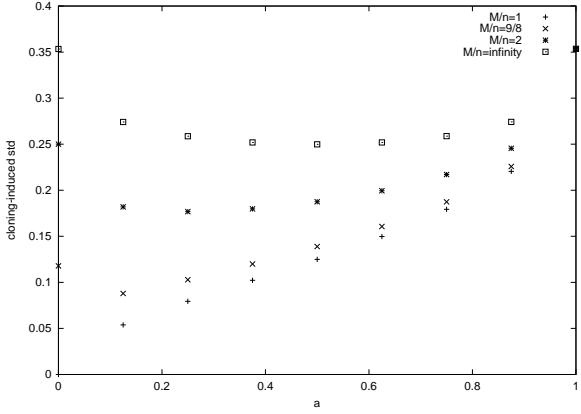}
\caption{Cloning-induced noise standard deviation $\sqrt{\mean{n_{th}}}$
as a function of the phase-conjugate fraction $a=N'/n$,
for $n=8$ and several values of $M/n$.}
\label{fig:gder}
\end{figure}

From Eq.~(\ref{amplifier}), we see that for fixed values of the total number of input replicas $n$ and number of output clones $M$, the gain $G$ (and thus the noise of the clones $\mean{n_{th}}$) only depends on $a$ and varies as
\begin{equation}
\label{eq:gr}
G(a)=\left(\frac{\sqrt{a} \sqrt{\frac{M}{n}+(2a-1)}
-\sqrt{\frac{M}{n}}\sqrt{1-a}}{2a-1}\right)^2
\end{equation}
In Fig.~\ref{fig:gder}, the value of $\sqrt{\mean{n_{th}}}$ is plotted as a function of $a$ for $n=8$ and different values of $M\geq n$. In the trivial case where $M=n=8$, the minimum additional variance is of course zero, and is obtained for $a=0$. The cloning transformation is then just the identity. However, when $M \geq n+1$, using phase-conjugated input modes yields lower variances than standard cloning if $a$ is correctly chosen (the lowest variance is then always attained for $a \neq 0$).  Remarkably, the value of $a$ achieving the minimum variance is not equal to $1/2$ for finite $M$, that is the optimal input partition contains more replicas than anti-replicas. In the limit of large $M$, however, the number of anti-replicas achieving the lowest variances tends to $n/2$, and the curve $G(a)$ tends to a symmetric curve around $a=1/2$. This behavior was expected, since $M =\infty$ corresponds to a measurement \cite{gisi97,cerf00:coherent} and we expect that measuring the value of $\psi$ from $N$ replicas of $\ket{\psi}$ and $N'$ replicas of $\ket{\psi^*}$ is equivalent to a cloning transformation starting from $N'$ replicas of $\ket{\psi}$ and $N$ replicas of $\ket{\psi^*}$. So, we conclude that the optimal measurement is achieved with balanced inputs ($N=N'$), as previously mentioned. Finally, in the case where $a=1$, the transformation consists in producing $M$ clones of $\ket{\psi}$ from $n$ replicas of $\ket{\psi^*}$. This is just phase-conjugation, for which we know that the best  strategy is to perform a measurement \cite{cerf01:pc}. The additional variance is therefore given by $1/n$, which does not depend on $M$. This explains why the curves converge all to the same point at $a=1$.

\section{Conclusions}

In summary,
we have studied continuous-variable cloning machines, which produce several
copies from one or more replicas of an arbitrary coherent state.  We have
derived the optimal fidelity of such cloners, as well as the actual cloning
transformations and the potential methods to implement them. We have then 
proposed a quantum key distribution protocol relying on continuous variables,
and shown how to apply reconcilation and privacy amplification to the
generated continuous key elements. We have investigated the balance between the information gained by the eavesdropper and that received by the authorized receiver, using cloning as an optimal individual eavesdropping strategy.
Finally, we have analyzed a new class of continuous-variable cloning machines,
which admit phase-conjugated inputs in addition to the normal inputs. 
By exploiting the antiunitarity of phase-conjugation, these new
cloners can beat the standard cloners in some cases. There is in general some
non-zero optimal phase-conjugate input fraction in order to 
maximize the cloning fidelity. As a conclusion,
it should be emphasized that these phase-conjugated inputs cloners 
do not extend on a qubit-based concept, in contrast with all previously
developed information-theoretic processes for continuous quantum variables.
Such a qubit cloner, admitting additional flipped qubits as inputs, 
has yet to be found.

We thank Fr\'ed\'eric Grosshans for comments on this manuscript, and Jean Cardinal and Serge Massar for useful discussions. N. J. C. is funded in part by the project EQUIP under the IST-FET-QJPC European programme. S. I. acknowledges support from the Belgian FRIA foundation. G. V. A.
acknowledges support from the {\it Communaut\'e Fran\c caise de Belgique} under an {\it Action de Recherche Concert\'ee}.

\appendix

\section{Density Matrices of Encoding Rules}
\label{sec:density}

In this Appendix, we would like to give further details regarding the protocol defined in section~\ref{sec:qdgk}. In particular, we will prove the equality of the density matrices $\rho_x$ and $\rho_p$ corresponding to Alice's two encoding rules provided that eq.~(\ref{eq:variances}) is  verified.

Define the Gaussian states $| \psi_x(r, \sigma_x) \rangle$ such that $\langle x \rangle = r$, $\langle p \rangle = 0$, $\Delta x^2 = \sigma_x^2$ and $\Delta p^2 = 1/4\sigma_x^2$. Similarly, let $| \psi_p(r, \sigma_p) \rangle$ be such that $\langle x \rangle = 0$, $\langle p \rangle = r$, $\Delta x^2 = 1/4\sigma_p^2$ and $\Delta p^2 = \sigma_p^2$. With the eigenstates $| x \rangle$ of $x$, our states have the following scalar products:
\begin{align}
\langle x | \psi_x(r, \sigma_x) \rangle &= \frac{1}{\sqrt{\sigma_x} \sqrt[4]{2\pi}} e^{-(x-r)^2/4\sigma_x^2} \\
\langle x | \psi_p(r, \sigma_p) \rangle &= \frac{\sqrt{2\sigma_p}} {\sqrt[4]{2\pi}} e^{-\sigma_p^2 x^2} e^{irx}
\end{align}
The density matrices $\rho_x$ and $\rho_p$ are defined as:
\begin{gather}
\rho_x = \int_{-\infty}^{+\infty} dr \frac{e^{-r^2/2\Sigma_x^2}}{\Sigma_x \sqrt{2\pi}} | \psi_x(r, \sigma_x) \rangle \langle \psi_x(r, \sigma_x) |
\\
\intertext{and}
\rho_p = \int_{-\infty}^{+\infty} dr \frac{e^{-r^2/2\Sigma_p^2}}{\Sigma_p \sqrt{2\pi}} | \psi_p(r, \sigma_p) \rangle \langle \psi_p(r, \sigma_p) | 
\end{gather}
Let us now calculate $\langle x | \rho_x | x' \rangle$ and $\langle x | \rho_p | x' \rangle$ in order to show that $\rho_x = \rho_p$.
\begin{equation}
\langle x | \rho_x | x' \rangle = \int_{-\infty}^{+\infty} dr \frac{e^{-r^2/2\Sigma_x^2-(x-r)^2/4\sigma_x^2-(x'-r)^2/4\sigma_x^2}}{\sigma_x \Sigma_x 2\pi} 
\end{equation}
The exponent of $e$ in the above equation can be rewritten as
\begin{equation}
\begin{split}
-\frac{(r-\frac{\Sigma_x^2 (x+x')}{2(\sigma_x^2 + \Sigma_x^2) })^2 }{2\Sigma_x^2 \sigma_x^2 / (\sigma_x^2 + \Sigma_x^2) }
&-\frac{x^2+{x'}^2 }{4(\sigma_x^2+\Sigma_x^2) }
\\
&-\frac{\Sigma_x^2 (x-x')^2 }{8\sigma_x^2 (\sigma_x^2+\Sigma_x^2) }.
\end{split}
\end{equation}
After integration, this yields
\begin{equation}
\langle x | \rho_x | x' \rangle = \frac{e^{-\frac{x^2+{x'}^2}{4(\sigma_x^2+\Sigma_x^2)}}e^{-\frac{\Sigma_x^2(x-x')^2}{8\sigma_x^2 (\sigma_x^2+\Sigma_x^2)}}}{\sqrt{2\pi} \sqrt{\sigma_x^2+\Sigma_x^2}}.
\end{equation}
For $\rho_p$, we have
\begin{equation}
\begin{split}
\langle x | \rho_p | x' \rangle &= \int_{-\infty}^{+\infty} dr \frac{2 \sigma_p}{2 \pi \Sigma_p} 
e^{-r^2/2\Sigma_p^2} \\
& \qquad \times e^{ir(x'-x)} e^{-\sigma^2_p(x^2+{x'}^2)} 
\\
&= \frac{2 \sigma_p}{\sqrt{2\pi}} e^{-\sigma^2_p(x^2+{x'}^2)} e^{-\frac{\Sigma_p^2}{2} (x-x')^2}.
\end{split}
\end{equation}

Taking (\ref{eq:variances}) into account, we have $\langle x | \rho_x | x' \rangle = \langle x | \rho_p | x' \rangle$ for all $x$, $x'$. Therefore, $\rho_x = \rho_p$.

\bibliography{qit}

\end{document}